# Modelling accretion in protobinary systems

Matthew R. Bate, Ian A. Bonnell and Nigel M. Price
*Institute of Astronomy, Madingley Road, Cambridge, CB3 0HA*

**ABSTRACT**

A method for following fragmentation simulations further in time using smoothed particle hydrodynamics (SPH) is presented. In a normal SPH simulation of the collapse and fragmentation of a molecular cloud, high-density regions of gas that form protostars are represented by many particles with small separations. These high-density regions require small time steps, limiting the time for which the simulation can be followed. Thus, the end result of the fragmentation can never be definitively ascertained, and comparisons between cloud fragmentation calculations and the observed characteristics of stellar systems cannot be made.

In this paper, each high-density region is replaced by a single, non-gaseous particle, with appropriate boundary conditions, which contains all the mass in the region and accretes any infalling mass. This enables the evolution of the cloud and the resulting protostars to be followed for many orbits or until most of the original cloud mass has been accreted.

The Boss & Bodenheimer standard isothermal test case for the fragmentation of an interstellar cloud is used as an example for the technique. It is found that the binary protostellar system that forms initially does not merge, but instead forms a multiple system. The collapse is followed to 4 initial cloud free-fall times when approximately 80 per cent of the original mass of the cloud has been accreted by the protostars, or surrounds them in discs, and the remainder of the material has been expelled out to the radius of the initial cloud by the binary.

## 1 INTRODUCTION

Recent observations of main-sequence stars show that most stars are in binary or multiple systems (Duquennoy & Mayor 1991). Pre-main-sequence stars also occur primarily in binary systems (Mathieu, Walter & Myers 1989; Simon et al. 1992; Ghez, Neugebauer & Matthews 1993; Leinert et al. 1993; Reipurth & Zinnecker 1993) and even show an excess of companions over the main sequence in most of the separation ranges surveyed. Any complete theory of star formation must therefore explain the formation of binary and multiple systems.

A theory for the formation of binary systems must account for the many observed properties of these systems. Duquennoy & Mayor (1991) found that binary stars have separations ranging from a few $R_\odot$ to $10^4$ au. The distribution of separations is unimodal with a broad peak at $\approx 30$ au. They also found that only the closest binaries have circular orbits, due to tidal circularization. The eccentricity of longer period binaries increases with period, and there is a lack of longer period systems with eccentricities of less than 0.15. Finally, in contrast with earlier studies, Duquennoy & Mayor (1991) found that most binaries have unequal masses with the mass ratio distribution increasing for smaller mass ratios.

Fragmentation is a possible mechanism to explain the formation of most binary systems. Most fragmentation calculations have concentrated on the isothermal collapse of low-mass molecular cloud cores. Early calculations involved the fragmentation of spherical clouds due to either an initial non-axisymmetric perturbation or the halting of the collapse due to rotation (Boss & Bodenheimer 1979; Bodenheimer, Tohline & Black 1980; Monaghan & Lattanzio 1986; Boss 1986). More recently, calculations have concentrated on the collapse of elongated clouds which fragment purely due to the initial shape of the cloud (Bonnell et al. 1991, 1992; Bonnell & Bastien 1992; Boss 1993; Nelson & Papaloizou 1993; Monaghan 1994). This type of fragmentation can form binary systems with separations from 10 to $10^4$ au. They have large eccentricities and a wide variety of mass ratios in agreement with observations.

Another type of fragmentation is the fragmentation of a protostellar disc. This has the potential to form binary systems with separations of the order of or less than a few hundred au. Two mechanisms have been investigated. The first is due to a pure disc process which excites an $m = 1$ mode moving the central object away from the system's centre of mass and allowing the disc to fragment (Adams, Ruden & Shu 1989; Shu et al. 1990; Heemskerk, Papaloizou & Savonije 1992; Woodward, Tohline & Hachisu 1994). The second generates an $m = 1$ mode from the interaction of a rapidly rotating core and continued infall on to the system with the disc (Bonnell 1994). This mechanism is also able to form close binary systems with separations of order 10 R$_\odot$ (Bonnell & Bate 1994).

All these calculations are forced to stop following the evolution soon after the formation of the fragments as the time steps required to follow the evolution accurately become too short. In calculations involving the isothermal collapse of molecular cores these fragments typically contain only a few per cent of the mass of the cloud. For the formation of close binaries, the situation is worse with less than 0.01 M$_\odot$ being contained in the fragments (Bonnell & Bate 1994). This makes it impossible to compare the results of fragmentation calculations with observations, since the final properties of the binaries will depend on the accretion of, or the interaction with, the rest of the cloud material. Indeed, the accretion of the remaining material may even force some seed binaries to merge.

In this paper, a method for extending SPH cloud fragmentation calculations by replacing protostars with non-gaseous, accreting particles is presented. This allows a collapsing cloud to be evolved beyond fragment formation to follow the infall of the rest of the cloud's material on to the seed protostellar system. Section 2 describes the SPH code and the inclusion of non-gaseous, accreting particles. A test of the boundary conditions for the non-gaseous, accreting particles is performed in Section 3. The evolution of the standard isothermal fragmentation test case proposed by Boss & Bodenheimer (1979), using both standard SPH and SPH with non-gaseous, accreting particles, is presented in Section 4. The conclusions are presented in Section 5.

## 2 CALCULATIONS

### 2.1 The SPH code

The calculations presented here were performed using a three-dimensional smoothed particle hydrodynamics (SPH) code based on a version developed by Benz (Benz 1990; Benz et al. 1990). The SPH technique can be considered as an interpolation scheme. The value of a physical quantity that varies in time and space can be estimated at a particular position by interpolating from a finite number of neighbouring positions where the values of the quantity are known. The interpolated value is calculated using a weighted sum of the known values over a kernel function, $W$. In SPH these interpolation points are particles that move with the fluid, making SPH a Lagrangian method. For example, the density of each particle is estimated by a weighted sum of the mass of the particle itself and the mass of each of its neighbours as

$$\rho_i = \sum_j m_j W_{ij}. \qquad (2.1)$$

The weighting, $W_{ij} = W(r_{ij}, h_{ij})$, for each particle depends on the separation of particles $i$ and $j$, $r_{ij} = |\mathbf{r}_i - \mathbf{r}_j|$, and their mean smoothing length, $h_{ij} = (h_i + h_j)/2$. The smoothing lengths of the particles give a distance scale over which particles interact. The smoothing lengths can vary in time and space, and do so to keep the number of neighbours approximately constant. A common form of kernel $W$, first suggested by Monaghan & Lattanzio (1985), is used. It is based on spline functions and is given by

$$W(r,h) = \frac{1}{\pi h^3} \begin{cases} 1 - \frac{3}{2}q^2 + \frac{3}{4}q^3 & \text{if } 0 \leq q < 1 \\ \frac{1}{4}(2-q)^3 & \text{if } 1 \leq q < 2 \\ 0 & \text{otherwise} \end{cases} \qquad (2.2)$$

where $q = r_{ij}/h_{ij}$ and the $i$ and $j$ subscripts have been dropped. The code uses a tree to calculate the gravitational forces and to find the nearest neighbours. The most common form for the artificial viscosity (Monaghan & Gingold 1983; Benz 1990; Monaghan 1992) is used, with the viscosity parameters $\alpha_v = 1$ and $\beta_v = 2$. The SPH equations are integrated using a second-order Runge-Kutta integrator.

In cloud fragmentation calculations the densities of gas frequently range over many orders of magnitude. High-density regions of gas are represented in SPH by many particles with small separations. These particles are limited by their separations and velocities to take shorter time steps than those in lower density regions. A great saving in computational time can be made by using individual particle time steps (Navarro & White 1993; Hernquist & Katz 1989). Each particle has its time step calculated and binned into time steps which differ by a power of two from the minimum time step allowed by the code. The time steps for each particle are limited by the Courant condition and a force condition (Monaghan 1992). The Courant condition gives a maximum allowable time step for a particle of

$$\delta t_{\rm cr} = \frac{0.3\, h}{c_{\rm s} + h|\nabla \cdot \mathbf{v}| + 1.2(\alpha_v c_{\rm s} + \beta_v h|\nabla \cdot \mathbf{v}|)}, \qquad (2.3)$$

where $c_s$ is the sound speed, $\mathbf{v}$ is the velocity of the particle, $\alpha_v$ and $\beta_v$ are the two artificial viscosity parameters, and the last term in the denominator is only included if $\nabla \cdot \mathbf{v} < 0$. The force condition gives a maximum time step of

$$\delta t_{\rm f} = 0.3 \sqrt{\frac{h}{|\mathbf{F}|}}, \qquad (2.4)$$

where $\mathbf{F}$ is the net acceleration on a particle. In addition, the use of the Runge-Kutta integrator allows a third time-step limit which is set by requiring that the changes in a particle's velocity, acceleration, internal energy and smoothing length, between two time steps, are less than some given tolerance. This is given by

$$\delta t_{\rm RK} = \sqrt{\frac{512\, \delta t_{\rm old}\, \lambda}{|Q_{\rm new} - Q_{\rm old}|}}, \qquad (2.5)$$

where $Q$ is the physical quantity, and $\lambda$ is the given tolerance. The minimum of these three values is used to set the time step for each particle.

The individual-time-steps code was tested against an earlier, single-time-step version. The single-time-step code requires only the Runge-Kutta integrator time-step limit. Shock tube tests, simulations of the adiabatic collapse of an initially isothermal, $1/r$ centrally condensed, spherical cloud (Evrard 1988; Hernquist & Katz 1989; Steinmetz & Müller 1992) and various other collapse calculations were performed. For the adiabatic collapse test, the results were also compared with a one-dimensional, Lagrangian, finite-difference code. The two SPH codes gave the same results, so long as both the Courant and force time-step conditions were included in the individual-time-steps code. The force condition was required in addition to the Courant condition in the individual-time-steps code as, without it, rapid changes in the forces were not followed quickly enough, resulting in the adiabatic collapse being too slow. The SPH adiabatic collapse test results also compared well to those of the finite-difference code, except for the effects of lower resolution (Steinmetz & Müller 1992).

In isothermal cloud fragmentation calculations, SPH simulations must be halted soon after a protostellar fragment forms due to the short time steps required for particles in the high-density fragments. This occurs despite the use of individual particle time steps. One method of following the calculations after the protostellar fragments form is to have a minimum particle smoothing length. This essentially changes the isothermal equation of state into that of an incompressible gas for the highest-density particles. Their separations then no longer decrease and the simulation can be followed longer, at the expense of ignoring the internal evolution of the protostars (e.g. Bonnell et al. 1991). A second method is to change the isothermal equation of state to a polytropic equation of state above a critical density (Bonnell 1994). This is motivated by the fact that, at densities of order $10^{-13}$ to $10^{-14}$ g cm$^{-3}$, the gas becomes optically thick. The collapse is no longer isothermal, but instead the gas is heated as it collapses. Following Bonnell (1994), the polytropic equation of state

$$P = K\rho^\gamma \tag{2.6}$$

is used, where $P$ is the pressure, $\rho$ is the density, and $K$ is a constant that depends purely on the entropy of the gas. The polytropic constant $\gamma$ varies with density as

$$\begin{aligned}\gamma &= 1.0, \qquad \rho \leq 10^{-14} \text{ g cm}^{-3}, \\ \gamma &= 1.4, \qquad \rho > 10^{-14} \text{ g cm}^{-3},\end{aligned} \tag{2.7}$$

and the gas has a temperature of 10K at $\rho = 10 \times 10^{-14}$ g cm$^{-3}$. This allows the calculations to be followed further as, in heating the gas, the collapse of high-density fragments is slowed. Indeed, with $\gamma = 1.4$ the collapse of a fragment proceeds only as its mass increases, since for a constant mass the thermal energy grows faster than does the magnitude of its gravitational energy. Despite this, however, the additional, unphysical change in $\gamma$ of

$$\gamma = \frac{5}{3}, \qquad \rho > 10^{-12} \text{ g cm}^{-3}, \tag{2.8}$$

is also required for the highest density regions to stop them collapsing under the increasing mass of the accreting fragments. This final change of the equation of state only occurs in the very centres of the fragments and is not as visible to particles outside the highest density regions as the minimum smoothing length method since the smoothing lengths are still allowed to decrease. Hence, the polytropic equation of state provides a more physically correct evolution than the minimum smoothing length method. Both methods maintain finite time steps for particles in fragments, and the calculation can be followed for much longer than with a purely isothermal equation of state.

## 2.2 The inclusion of non-gaseous, accreting particles in SPH

Calculations done with a minimum smoothing length or a polytropic equation of state are still very computationally expensive, however, and become slower as more particles move to the higher density regions and hence have small time steps. Since the internal evolution of the protostars is not of interest in determining where material from the cloud ends up, a protostar is replaced by a single, non-gaseous, massive particle with the combined mass, linear momentum and 'spin' of the particles it replaces. This massive particle then accretes any infalling gas particles and avoids the problem of high-density regions controlling the time required for the simulation. The spin is the angular momentum of the replaced particles about their centre of mass and is kept as a check that global angular momentum is being conserved, although it has no effect on the dynamics of the calculation.

This idea has been used before (Boss & Black 1982; Boss 1987, 1989) to study the collapse of rotating, isothermal interstellar clouds. In these papers, a finite-difference code was used where the central cell was designated as a 'sink cell' which could accrete any infalling gas and its angular momentum. In this way, they were able to study the collapse of clouds beyond the formation of a central protostellar core. These sink cells were fixed in position, however, and therefore accretion on to a binary or multiple protostellar system cannot be studied using this method.

By contrast, the inclusion of dynamic, non-gaseous, accreting particles (sink particles) into SPH is relatively simple, due to the Lagrangian nature of SPH. It involves three main parts. First, the method of accreting infalling gas particles must be decided. Secondly, a method for the dynamic creation of a sink particle when a protostellar fragment is created during a collapse calculation is required. Finally, boundary conditions between the sink particles and the surrounding SPH gas particles are required.

### 2.2.1 *Accretion by sink particles*

The sink particle itself is simply a particle, tagged to identify it as a sink particle, with a mass much greater than a typical gas particle (i.e. greater than $\approx 50$ times). Sink particles only interact with normal gas particles via gravity (and boundary conditions as will be discussed later). Also, gas particles do not interact with other gas particles on the opposite side of a sink particle, except via gravity. Particles that come within a certain radius, the accretion radius $r_{\text{acc}}$, of the sink particle may be accreted. The accretion radius is chosen before the calculation is performed, depending on the resolution required near the sink particle, and remains fixed throughout. All gas particles that come within the accretion radius pass through several criteria to see if they should be accreted. First, the particle must be bound to the sink particle. Secondly, the specific angular momentum of the particle about the sink particle must be less than that required for it to form a circular orbit at $r_{\rm acc}$ about the sink particle. These conditions ensure that particles which would normally leave the accretion radius are not accreted. Finally, the particle must be more tightly bound to the candidate sink particle than to any other sink particle. If two sink particles are of unequal mass, it is possible that a particle may briefly pass through the accretion radius of the least massive sink particle, but finally be accreted by the more massive sink particle. This test is included to stop particles being prematurely accreted by the wrong sink particle. Care must also be taken in an individual-particle-time-steps code to ensure that the particle being accreted has just completed a time step.

In addition to the accretion radius described above, a second, inner accretion radius is commonly used. All gas particles coming within this inner radius (which is set to be $10-100$ times smaller than the outer accretion radius) are accreted regardless of the tests. This stops particles being accelerated by a close encounter with the sink particle. An alternative is to smooth the potential of the sink particle, as is done in N-body calculations. This inner accretion radius will not be mentioned further.

When a gas particle is accreted, its mass and linear momentum are added to the sink particle. Also, the sink particle's position is shifted to occupy the centre of mass of the gas particle and sink particle. Finally, the spin of the sink particle is modified to take account of the angular momentum about the sink particle that is accreted. The gas particle is then tagged as being accreted, and is no longer involved in the calculation.

### 2.2.2 Dynamic Creation of Sink Particles

Sink particles can be either included in a simulation from the beginning, or dynamically created during a cloud collapse calculation. The procedure for dynamically creating a sink particle is as follows. When a gas particle in an SPH simulation reaches a specified density (say $10^5$ times the initial cloud density), a series of tests are performed to determine if a sink particle should be created from it (see also Price, Bate & Bonnell (1995) for an analytic determination of the density at which testing for sink particle creation should begin). It is first tested to see if its smoothing length is less than half the accretion radius of the sink particle that will replace it. This is done to ensure that a sink particle is formed from at least $\approx 50$ particles (the mean number of neighbours). If this is so, and all the sink particle's neighbours are being evaluated at the current time step (particles are binned into time steps differing by a power of two), then, at the end of this step, the particle and its neighbours undergo a series of tests to decide if they should create a sink particle.

The gas particles should only form a sink particle if they will continue to collapse, if not replaced by a sink particle, rather than expand again. To ensure this, the gas particles must pass four tests. First,

$$\alpha \leq \frac{1}{2}, \tag{2.9}$$

where $\alpha$ is the ratio of thermal energy to the magnitude of the gravitational energy for the particles. Secondly,

$$\alpha + \beta \leq 1, \tag{2.10}$$

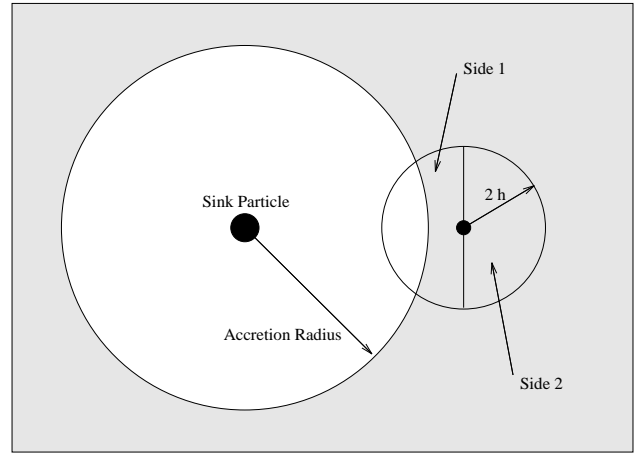

**Figure 1.** The configuration of a typical SPH gas particle near the accretion radius of a sink particle, demonstrating the need for boundary conditions. The volume that contains SPH gas particles is shaded, whereas the inside of the accretion radius contains no gas particles. Hence, the volume of the kernel of a gas particle (of a radius of twice the smoothing length, $h$) near the boundary has some of its neighbours missing. When the density (or another quantity) is calculated for that particle, the missing neighbours affect the contribution to the density from side 1 of the gas particle, but the contribution from side 2 is unaffected by the accretion hole.

where $\beta$ is the ratio of rotational energy to the magnitude of the gravitational energy for the particles. Rotation is calculated about the position of the densest particle. Thirdly, the total energy of the particles must be negative. Finally, the divergence of the accelerations on the particles must be less than zero. This final check is done to determine if the particles for the candidate sink particle have a net acceleration away from one another. If they do, the dense region could be in the process of being tidally disrupted or bouncing and should not form a sink particle. In test calculations, sink particle creation almost never fails on this final test alone. If all the tests are passed, a sink particle is formed from the densest gas particle by combining it with its neighbouring gas particles and moving it to the centre of mass.

### 2.2.3 Boundary condition for sink particles

Ideally, the introduction of sink particles to an SPH simulation should not affect the evolution of the gas outside the accretion radius. In practice, however, two effects will alter the evolution. First, there will be some effect because the typically non-spherical, gravitational potential is replaced by a spherical point mass potential, though this should generally be small. Secondly, and most importantly, there is the discontinuity in the number of particles across the accretion radius due to particles inside being accreted. This affects the pressure and viscous forces on particles outside. Boundary conditions for gas particles near the accretion radius are necessary to correct for this.

Four types of correction to particles near the accretion radius are required. First, the smoothing lengths of particles near the boundary will be larger than they would otherwise be, as they are still required to have ≈ 50 neighbours, but they are missing neighbours inside the accretion radius (see Fig. 1). Second, the density of particles near the boundary is underestimated because the summation in equation (2.1) is no longer over the entire kernel due to the missing neighbours. Thirdly, particles near the boundary will tend to be pushed into the accretion hole because of a lack of pressure forces from the missing particles. Finally, the standard SPH artificial viscosity has bulk viscosity and shear viscosity components (Meglicki, Wickramasinghe & Bicknell 1993). Although in the continuum limit ($h \to 0$) the viscosity present in SPH has no shear viscosity, in a real SPH simulation shear viscosity is present due to the interaction of particles with each other across a shear flow over a finite distance of the order of $h$. If viscous boundary corrections are not made, the lack of neighbours inside the accretion radius can lead to unrealistic angular momentum transport for particles near the boundary via this shear viscosity. For example, in an accretion disc simulation, particles near the boundary are quickly accreted by the sink particle as they lose angular momentum and fall inside the accretion radius (see Section 3). They still have all of their neighbours further out in the disc which are rotating less rapidly about the sink particle and hence removing angular momentum from them via the shear viscosity. However, they are missing neighbours closer to the sink particle that would have given them angular momentum. Thus, they lose angular momentum much more quickly than is correct, and are accreted, and a hole in the disc is rapidly cleared out around the sink particle.

The situations for which boundary conditions around sink particles may be neglected without seriously affecting the calculation are limited. Pressure-related boundary conditions are not required when the radial infall velocity on to the sink particle is supersonic, as this constrains the density and pressure effects to remain near the accretion radius. Viscosity-related boundary conditions are not required when the rotational velocity of the gas around the sink particle is very low compared with the radial infall velocity, or when the dynamics of the simulation change quickly compared with the viscous time-scale of the gas around the sink particle (e.g. due to gravitational torques). In these cases, the effect of the angular momentum transport due to the shear viscosity is small. Note, however, that the viscous time scale is a lot shorter at the boundary than it would be if the medium were continuous, due to the large density gradient across the accretion radius. Examples of cases where boundary conditions are not required are spherical, zero-angular-momentum accretion, and Bondi–Hoyle accretion. Simulations of both these types of accretion have been performed by Price et al. (1995) using sink particles without boundary conditions. They were found to reproduce one-dimensional, finite-difference code and analytic results well. In most other cases, however, boundary conditions are required.

To avoid the discontinuities at the boundary, we experimented with many different types of boundary condition. One method was to evolve particles, once they had entered the accretion radius, purely under gravity, maintaining constant inward radial velocity, or constant angular momentum, or the rate of angular momentum loss. This method did not provide good viscous boundaries and typically slowed down the calculations by factor of two or three as particles inside the accretion radius were evolved every minimum time step along with the real particles. For the pressure boundary, another attempted method was to add a radial pressure force to particles near the accretion radius from the sink particle by giving the sink particle an effective mass, calculated from its accretion radius and a density extrapolated from its neighbouring particles' densities. This method did not give consistent results for different sized accretion radii.

The final boundary conditions arrived at are as follows. To correct the density and smoothing lengths of particles near the boundary, a linear approximation to the density gradient at each particle's position is found. The neighbours of a particle are sorted by radius from the sink particle and split into two groups. The mean density and radius of these two groups of particles are calculated, allowing a linear fit to the density to be found. The densities from the previous time step are used. This provides an estimate of the radial density gradient at each particle position. In the calculation for a particle's density at the current time step, the contributions to its density in equation (2.1) from its neighbours on the side facing the sink particle (side 1), and the other side (side 2) are calculated (see Fig. 1). The former is affected by the accretion hole whereas the latter is not. Half of the self-contribution to the particle's density is also added to contributions from each side of the particle. By integrating the analytic equation for the density,

$$\rho = \int \rho W \, \mathrm{d}^3 r \qquad (2.11)$$

over the volume of the kernel $W$, assuming a linear density gradient $\phi$, the expected contribution to the particle's density from each side can be determined. The density is corrected by making the contribution to the density from side 1 of the particle's kernel be the maximum of the SPH summation value, and the value given by

$$\rho_1 = \rho_2 \kappa, \qquad (2.12)$$

where the subscripts denote sides 1 and 2, and $\kappa$ is the ratio of the analytic contribution of side 1 to the particle's density to side 2's contribution, assuming the linear density gradient. The ratio $\kappa$ is given by

$$\kappa = \frac{1-c}{1+c}, \quad \text{where} \quad c = \frac{31 h \phi}{70 \rho} \qquad (2.13)$$

when the kernel of equation (2.2) is used. This ratio is also used to find an approximation to the missing number of neighbours of the particle. Assuming the neighbours are roughly equally spaced on both sides, $\kappa$ gives the ratio of the numbers of neighbours on each side of the particle, allowing the number of missing neighbours to be determined. This extra number of missing neighbours is taken into account when altering the particle's smoothing length.

For the pressure correction, only radial pressure accelerations on boundary particles are corrected. Using the same method as is used to determine the density gradient, an approximation to the pressure gradient at a particle's position is determined. Then using

$$\frac{\mathrm{d}_\mathrm{p}}{\mathrm{d}t} = -\frac{1}{\rho} \frac{\mathrm{d}P}{\mathrm{d}r}, \qquad (2.14)$$

the expected radial acceleration due to pressure on the particle is found. This is compared with the radial pressure acceleration on the particle as determined by the SPH summation equations. If the expected value gives a greater outward acceleration than the SPH value, a radial acceleration correction, which is the difference of the expected and SPH values, is added to the particle's acceleration. To conserve linear momentum, an equal and opposite force is applied to the sink particle.

Boundary conditions for the artificial viscosity forces are not as simple as the other corrections. With shear viscosity, forces can be given to a particle from shear flows in several directions. Also, the magnitude of the force depends on the second derivative of the velocity across the shear flow, and on the density gradient. In addition, viscous forces on particles can be given by shocks as well as shear flow. Analytic corrections become very complicated.

Instead, a simpler method is used. First, the direction of the smoothed velocity is calculated for each particle near the boundary and for all their neighbours. This is calculated by summing the particle's velocity and all its neighbours' velocities, weighted by the kernel. Then, for each particle, the components of these smoothed velocities that are orthogonal to the radial vector to the sink particle are found. Using these velocities, the component of the viscous acceleration on a particle that is directed tangentially to the accretion hole, and in the direction of motion of the particle, is found for each particle using the corrected viscous accelerations from the previous time step. Once this component of the viscous acceleration has been calculated for each particle, an estimate of the correct viscous acceleration on a particle in this direction is calculated as the mean of these values for each of the particle's neighbours that are further in radius from the sink particle than the particle itself. Using the estimate of the viscous tangential acceleration in this direction, a correction is added to the particle's viscous acceleration in a similar way to the radial pressure acceleration correction. The value of the correction is the difference between the estimated viscous tangential acceleration and the SPH value calculated at the current time step in this direction, with the constraint that the net viscous acceleration on the particle cannot be in its direction of motion to avoid giving particles extra angular momentum (i.e. particles near the boundary can be stopped from losing angular momentum, but cannot gain angular momentum). Again, to conserve linear momentum an equal and opposite force is applied to the sink particle, and in this case a correction to the sink particle's spin is also made to conserve angular momentum. If the constraint that a particle near the boundary cannot increase its angular momentum is not included, two effects can lead to incorrect results. First, a runaway situation is possible. The particle may interact with its neighbours that are further from the sink particle, giving them accelerations in their direction of motion, in turn increasing the acceleration given to the boundary particle. Secondly, the correction is intended to handle the shear viscosity discontinuity at the boundary, but viscous accelerations can also be given by shocks, which may not be estimated well. Since the angular momentum for the correction comes from the sink particle's spin there is, potentially, an unlimited source of angular momentum for these two effects to inject into the surrounding gas. Hence, allowing angular-momentum-increasing viscous accelerations can lead to incorrect results and they should not be allowed.

Viscous accelerations on particles near the boundary that are directed radially, and viscous accelerations directed orthogonally to the direction of motion of the particle and the radial direction, are not altered even though they are also affected by the presence of the accretion hole. However, since the action of viscosity is to resist motion, and viscous accelerations in the direction of motion of the particles and orthogonal to the radial direction are corrected, the major boundary effect is corrected. To attempt to correct the viscous forces in the other directions by this method could be worse than having no correction at all, due to the presence of shocks.

Finally, the boundary conditions for a particle are applied gradually as the sink particle is approached. This is achieved by multiplying the corrections by a smoothly varying function $\chi$ ranging between zero and unity as the particle approaches the sink particle. The functional form of $\chi$ is relatively unimportant and for simplicity the unnormalized kernel function

$$\chi = \begin{cases} 1 & \text{if } x < 0, \\ \pi h^3 W(x,h) & \text{if } 0 \leq x \leq 2, \\ 0 & \text{if } x > 2, \end{cases} \quad (2.15)$$

is used, where $x = 2(r - r_{\rm acc} - h)/h$ for the corrections to the density and number of neighbouring particles, and $x = 4(r - r_{\rm acc} - 3h/2)/h$ for the corrections to pressure and viscous forces. The different forms of $x$ reflect the radii at which neighbouring particles have a large contribution to the calculation of a particle's density and the pressure and viscous forces upon it. In addition, if a particle is a neighbour of two or more sink particles, then no boundary corrections are applied to it.

A comparison of the quantities near the edge of an accretion radius between calculations with and without boundary conditions is shown in Fig. 2. The simulation has been created by placing a sink particle at the centre of an already-formed disc. The difference made by including boundary conditions is readily apparent. The smoothing lengths of particles near the boundaries no longer increase due to a lack of neighbours. The density does not drop off, but continues to increase toward the centre. The radial pressure accelerations on particles near the boundary are no longer inward, but are corrected to have net outward accelerations due to pressure. Finally, the mean of the viscous tangential accelerations (calculated for each particle from the viscous tangential accelerations on a particle and all its neighbours) are now approximately zero, rather than being negative, which would cause the particles to fall inside and be accreted. The mean over a particle and its neighbours is plotted rather than the actual viscous accelerations as, at any one moment, these tend to be very noisy. As previously mentioned, individual particles with boundary corrections are not allowed angular-momentum-increasing accelerations. This gives the slight drop of the mean viscous tangential accelerations in Fig. 2 toward the boundary and results in a slightly higher accretion rate of particles than if the full correction was used. Tangential viscous accelerations that give net angular momentum to a particle near the boundary can be allowed in a stable disc simulation, although, if shocks and other instabilities are present, it is safest not to allow angular-momentum-increasing corrections to be applied to particles.

mean of the viscous accelerations on particles farther out. In Bondi-Hoyle accretion, particles swing past the point mass, collide behind it, and then fall in toward the point mass and are accreted. If the smoothing lengths are too large, the viscous accelerations needed to stop the particles in the colliding streams are smoothed out too much by the process of taking the mean acceleration, and the shock behind the point mass is not resolved. This leads to less dissipation of energy, and too few particles are accreted. Note that only the viscous boundary conditions, and not the other three corrections, cause the disruption to Bondi-Hoyle accretion. This effect must be considered when setting the accretion radius. Alternatively, it may be possible to implement a switch, employing boundary conditions only when the net rotation of the gas around the sink particle is significant. For example, in an accretion disc, viscous boundary conditions would be used, but would not be necessary for Bondi-Hoyle accretion and spherical accretion.

## 3 DISC FORMATION AROUND A SINK PARTICLE

To demonstrate the effectiveness of the sink particle boundaries, results are presented from a calculation of the non-self-gravitating collapse of a spherical, rotating, uniform density cloud to form a disc around a sink particle. The initial conditions consist of a spherical, uniform cloud of molecular hydrogen in solid-body rotation with mass $M$, radius $R$, temperature $T$, and rotational frequency $\Omega$, of

$$M = 0.001 \text{ M}_\odot, \quad R = 2.56 \times 10^{16} \text{ cm},$$
$$T = 10 \text{ K}, \quad \Omega = 1.54 \times 10^{-12} \text{ rad s}^{-1}, \quad (3.1)$$

surrounding a 1-$M_\odot$ star at the centre of the cloud, which is represented by a sink particle. The cloud collapses isothermally, under the gravity of the central sink particle only, to form a non-self-gravitating disc around the sink particle. The calculation is performed with two different accretion radii. If the use of sink particles affects the dynamics of the gas outside the accretion radius, the mass contained within a given radius at any one moment will be different for simulations using different accretion radii.

**Figure 2.** The smoothing lengths (top), densities (upper middle), radial pressure accelerations (lower middle), and mean viscous accelerations in the direction of motion (bottom) of particles in a disc surrounding a sink particle. These are shown without (left) and with (right) boundary conditions at the accretion radius. The units are arbitrary with the gravitational constant $G = 1$.

Finally, zero-angular-momentum, spherical accretion and Bondi–Hoyle accretion calculations have been performed using sink particles with these boundary conditions. As with the calculations performed by Price et al. (1995), using sink particles without boundary conditions, the zero-angular-momentum, spherical accretion reproduced the one-dimensional, finite-difference results well. With Bondi-Hoyle accretion, the correct accretion rate is given provided that

$$h_{\text{acc}} \leq (r_{\text{BH}} - r_{\text{acc}}), \quad (2.16)$$

where $h_{\text{acc}}$ is the smoothing length of particles near the accretion radius, and $r_{\text{BH}}$ is the Bondi-Hoyle accretion radius. If $h_{\text{acc}}$ is larger than this, the accretion rate is lower than expected. This occurs because the viscous accelerations on particles near the accretion radius are corrected to be the

Fig. 3 shows the formation of the disc around the sink particle for four different calculations. The first two are performed with no boundary conditions at the accretion radius of the sink particle, and have $r_{\text{acc}} = 0.01$ and $r_{\text{acc}} = 0.05$ in units of $10^{16}$ cm. In both cases, the central region of the disc is cleared out to a radius much larger than the accretion radius, mainly due to the lack of a viscous boundary condition. When boundary conditions are added to the sink particle implementation, the situation is vastly improved. The early formation of the disc is much better handled, and later in the calculation the central regions of the discs are no longer cleared out due to the discontinuity. Instead, particles are present right up to the accretion radius. Near the end of the calculation, particles are still deficient at the centre of the disc, especially in the small-accretion-radius case. However, most of this is not due to the lack of a boundary condition. To understand this effect, we must consider how the mass contained within $r = 0.05$ of the sink particle changes with time. This is shown in Fig. 4. For the two calculations without boundary conditions, the rate at which material enters $r = 0.05$ is larger with the larger accretion radius than with the smaller accretion radius as the disc is formed. However, later on both calculations have the same growth rate of mass within $r = 0.05$. As is shown in Fig. 3, this is because the central regions of the disc are cleared out in both cases, faster initially with the larger accretion radius, but later both discs are cleared out to much greater than $r = 0.05$ and the accretion then depends on the flow of mass through the accretion disc, which is similar for both cases. When boundary conditions are included in the calculations, the masses contained within $r = 0.05$ of the sink particle differ even more, with less mass being contained by the smaller accretion radius calculation. However, the total mass of material that has passed within $r = 0.05$ of the sink particle, at some point during the calculation, is largely independent of accretion radius. This demonstrates that the boundary conditions of the sink particle are working well. The reason that the mass contained within $r = 0.05$ of the small-accretion-radius sink particle at a specific time is less than the mass that has been within this radius at any time during the calculation is that the evolution inside the radius is such that the outer material gains angular momentum from material further in due to shear viscosity and so moves out of the $r = 0.05$ region. This can be seen in the way in which the mass within $r = 0.05$ for the small accretion radius calculation decreases with time after the disc has formed. The deficiency of particles in the inner parts of the discs at late times seen in Fig. 3 is now explained.

**Figure 3.** The formation of a disc around a sink particle for four different cases. The first two cases have no boundary conditions at the accretion radius $r_{\rm acc}$, of the sink particle and have $r_{\rm acc} = 0.01$ (top) and $r_{\rm acc} = 0.05$ (upper middle). The second two cases are performed with boundary conditions at the accretion radius of the sink particle and again have $r_{\rm acc} = 0.01$ (lower middle) and $r_{\rm acc} = 0.05$ (bottom). The boundary conditions stop the excess accretion from the disc that occurs in calculations without boundary conditions due to the accretion hole. The calculation is performed with 5000 particles. The time, in units of the orbital period of a particle at $r = 0.05$, is given in the upper right hand corner of each panel. The axes are in units of $1.0 \times 10^{16}$ cm.

**Figure 4.** The formation of a disc around a sink particle, showing how the mass contained within $r = 0.05 \times 10^{16}$ cm of the sink particle (including the sink particle's original mass) increases with time. The results for an accretion radius of $r_{\rm acc} = 0.01 \times 10^{16}$ cm are shown with (solid) and without (short-dashed) boundary conditions. Calculations with $r_{\rm acc} = 0.05 \times 10^{16}$ cm are also shown with (dotted) and without (long-dashed) boundary conditions. Finally, the mass that has been inside $r = 0.05 \times 10^{16}$ cm at any time during the calculation with $r_{\rm acc} = 0.01 \times 10^{16}$ cm is shown for the case with boundary conditions (dot-dashed). The calculation was performed with 5000 particles. Mass is given in M$_\odot$, and the time is in the units of the orbital period of a particle at $r = 0.05 \times 10^{16}$ cm.

The boundary conditions are behaving well, but outward evolution of mass from the inner parts of the disc cannot be followed correctly since material that has been accreted cannot re-enter the simulation. Particles that would have evolved outward into the deficient zone were accreted earlier by the sink particle. A difference between the amount of mass within $r = 0.05$ at a specific time and with the mass that has been within this radius at any time did not occur in the small-accretion-radius calculation without boundary conditions, as accretion due to the lack of boundary conditions overwhelmed the gain in angular momentum from the material inside $r = 0.05$.

The transport outward also explains most of the difference between the large-accretion-radius and small-accretion-radius calculations when considering the mass that has been within $r = 0.05$ at some point during the calculation. When the disc is forming, the small-accretion-radius calculation transfers some angular momentum from inside $r = 0.05$ to outside via viscosity, allowing less material to enter the $r = 0.05$ region. This cannot occur with the large-accretion-radius calculation as there are no particles inside $r = 0.05$ and angular-momentum-increasing accelerations on particles near the boundary are specifically not allowed by the boundary conditions. Later, when the disc has fully formed, material enters $r = 0.05$ only slightly faster with the large accretion radius than with the small accretion radius. Again this is because angular-momentum-increasing accelerations are not allowed by the boundary conditions. Although the mean tangential accelerations on the particles near the boundary are nearly zero once the disc has formed, numerical noise means that some particles have negative and some positive accelerations, and, since positive accelerations are not allowed, the mean is slightly negative giving a slightly increased accretion rate. If angular-momentum-increasing accelerations are allowed by the boundary conditions, the small and large accretion radius calculations give more similar results. This is possible in a stable disc calculation, but, as was explained earlier, is not advisable generally. Despite this constraint, however, the boundary conditions give a vast improvement over the case without boundary conditions.

## 4 STANDARD ISOTHERMAL TEST CASE

We now demonstrate the promise of incorporating sink particles in SPH to allow the evolution of cloud fragmentation calculations beyond protostellar creation, and thus determine the final system states. The chosen example is the standard isothermal test case for the fragmentation of a molecular cloud. This was first described by Boss & Bodenheimer (1979), and since then has become a typical test case for numerical codes studying fragmentation processes (Bodenheimer et al. 1980; Bodenheimer & Boss 1981; Monaghan & Lattanzio 1986; Boss 1986; Myhill & Boss 1993). The initial conditions consist of an isothermal, spherical cloud of molecular hydrogen gas. The cloud's mass $M$, radius $R$, temperature $T$, and rotational frequency $\Omega$ are

$$M = 1.0 \text{ M}_\odot, \qquad R = 3.2 \times 10^{16} \text{ cm},$$
$$T = 10 \text{ K}, \qquad \Omega = 1.56 \times 10^{-12} \text{ rad s}^{-1}, \qquad (4.1)$$

resulting in ratios of thermal energy to the magnitude of gravitational energy, and rotational energy to the magnitude of gravitational energy of $\alpha = 0.25$ and $\beta = 0.20$, respectively. Note that a value of $\Omega$ slightly less than the value of $1.6 \times 10^{-12}$ rad s$^{-1}$ that is usually quoted is used, as this gives $\beta$ closer to 0.20. In addition, there is an $m = 2$ density perturbation imposed on the cloud depending only on the azimuthal angle $\phi$, given by

$$\rho = \rho_0[1 + 0.5 \sin(2\phi)] \qquad (4.2)$$

where $\rho_0 = 1.44 \times 10^{-17}$ g cm$^{-3}$. The initial cloud free-fall time is $t_{\rm ff} = 5.52 \times 10^{11}$ s.

Previous calculations of the standard test case have shown that the cloud forms two fragments due to the $m = 2$ density perturbation. Until now the fragmentation of this test case has not been followed beyond 1.9 free-fall times (Monaghan & Lattanzio 1986), and commonly is only evolved until about 1.3 free-fall times (Boss & Bodenheimer 1979; Bodenheimer et al. 1980; Bodenheimer & Boss 1981; Boss 1986; Myhill & Boss 1993). This is not long enough even to determine if the binary fragmentation survives, or merges as the two fragments fall toward each other. Indeed, it has been speculated that the fragments merge, and do not form a binary system (Monaghan & Lattanzio 1986).

### 4.1 Standard SPH fragmentation

To test how the evolution of a fragmentation calculation is affected by the use of sink particles, the calculation must first be evolved without sink particles. With a purely isothermal equation of state, the calculations are forced to halt at 1.27 initial cloud free-fall times($t_{\rm ff}$, due to the high densities reached. However, the introduction of a minimum particle smoothing length or a polytropic equation of state allows the standard test case to be evolved further. In two calculations, the standard test case was evolved with standard SPH to $2.2t_{\rm ff}$. One simulation maintained finite time steps via a minimum particle smoothing length, keeping the gas isothermal. The minimum smoothing length was set to be 1 per cent of the initial cloud radius. The other simulation used the polytropic equation of state described in Section 2.1 and no minimum smoothing length. Fig. 5 shows the fragmentation as calculated with the polytropic equation of state. The initial evolution is almost identical in both cases. The cloud initially forms two fragments that collapse on themselves. The fragmentation does not proceed via the initial formation of a bar structure, as seen in early calculations (Bodenheimer & Boss 1981). Instead, the two fragments are distinct from the beginning of their formation. This agrees with the later finite-difference code results (Myhill & Boss 1993).

**Figure 5.** The particle positions projected onto the equatorial plane for the Boss & Bodenheimer (1979) standard test case evolved with a polytropic equation of state using standard SPH. In past calculations, the standard test case is typically only evolved to $1.3t_{\rm ff}$, however, the use of a polytropic equation of state allows it to be followed further. Collapse of the cloud initially forms two protostars. These two protostars fall toward each other, forming a binary system due to their orbital angular momentum. Subsequently, the fragmentation of the disc surrounding one of the protostars forms a third protostar and a triple system results. The simulation was performed with 8024 particles. The time, in units of the free-fall time ($t_{\rm ff} = 5.52 \times 10^{11}$ s), is given in the upper right hand corner of each panel. The axes are in units of $1.0 \times 10^{16}$ cm.

**Table 1.** Fragment properties from the Boss & Bodenheimer (1979) standard test case performed with SPH at 1.32 initial cloud free-fall times. Columns M64 and B64 give the results of Myhill & Boss (1993) for comparison.

| Model | Isothermal | Polytropic | M64 | B64 |
|---|---|---|---|---|
| $M_{\rm f}/{\rm M}_\odot$ | 0.17 | 0.17 | 0.15 | 0.10 |
| $\alpha_{\rm f}$ | 0.02 | 0.10 | 0.08 | 0.13 |
| $\beta_{\rm f}$ | 0.24 | 0.28 | 0.29 | 0.38 |
| $(J/M)_{\rm i}/(J/M)_{\rm f}$ | 12 | 12 | 10 | 12 |
| $D/R$ | 0.29 | 0.29 | 0.28 | 0.32 |

To allow comparison of the SPH results with the results of Myhill & Boss (1993), the initial fragment properties are listed in Table 1 for both simulations at $1.32t_{\rm ff}$. The results of Myhill & Boss (1993) were obtained using Eulerian grid codes based on Cartesian (M64) and spherical (B64) coordinate grids. At this stage, the maximum density is $\sim 10^6 \rho_0$. The results are averaged over the two fragments. Fragment mass is calculated from all the particles in one half of the cloud with a density of $100\rho_0$ or greater. The quantities $\alpha_{\rm f}$ and $\beta_{\rm f}$ are then calculated from these particles. The fragment is assumed to be spinning parallel to the $z$-axis about its centre of mass, rather than about the most dense particle. The specific angular momentum of the fragment $(J/M)_{\rm f}$ is given by the angular momentum of the fragment about the centre of mass divided by the fragment mass. As in Myhill & Boss (1993) this is compared with the specific angular momentum of the initial cloud $(J/M)_{\rm i}$, as a measure of how the initial angular momentum of the cloud is partitioned between the spin and orbital angular momenta of the fragments. Finally, the distance from the fragment's centre of mass to the $z$-axis $D$, divided by the cloud radius $R$, is given.

The results are in good agreement with those of Myhill & Boss (1993). The polytropic equation of state does not affect the tabulated fragment properties significantly over the isothermal case (with the exception of $\alpha_{\rm f}$, due to heating of the high-density gas) as the equation of state only becomes non-isothermal beyond $1.18 t_{\rm ff}$.

Following the initial fragmentation of the cloud in Fig. 5, the fragments (or protostars) continue to fall toward each other due to their lack of angular momentum. However, they do not merge. Instead, the two initial protostars pass each other at periastron at approximately $1.9 t_{\rm ff}$ and continue to orbit each other. As demonstrated in Table 1, the protostars have a high ratio of rotational to gravitational energy. They quickly become bar-unstable, forming spiral arms that move through their discs (Fig. 5). Before periastron is reached, at $1.79 t_{\rm ff}$, the disc around one of the protostars fragments. The cause of this disc fragmentation is the same process as presented by Bonnell (1994). The protostar embedded in the disc is rapidly rotating, making it unstable to the growth of an $m = 2$ bar-mode. In forming a bar, the protostar is able transfer angular momentum to the outer parts of the bar via gravitational torques. These move outward and fall behind the bar, forming spiral arms. Any asymmetry in the two ends of the bar, or their spiral arms, leads to a different rate of angular momentum loss, and hence support, from each side of the protostar. The protostar then contracts non-uniformly, moving its centre of mass away from the centre of mass of the system and spiralling outward. This allows the spiral arms to catch up to one another and they collide. With the mass swept up by the spiral arms travelling through the disc, and the continued infall of material, the collision forms a gravitationally unstable condensation which collapses, forming a secondary. Hence, a triple system is formed.

Although the initial cloud is perfectly symmetric, only one disc has fragmented by $2.2 t_{\rm ff}$. This demonstrates that disc fragmentation is a highly chaotic process. Small variations in the particles' positions and velocities in the simulations come from the limited numerical precision of the calculations. These lead to a small amount of asymmetry which allows one disc to fragment before the other. It is highly probable that the gravitational interaction between the two protostars and their discs increases this asymmetry and aids the fragmentation (as suggested in Bonnell et al. 1992). Furthermore, once a close binary is formed, gravitational torques exerted by it on the other disc could help stabilize this disc against fragmentation. Thus, although it is possible to predict that fragmentation is likely to occur from the high ratio of rotational energy to the magnitude of the gravitational energy in the disc, the exact timing and position of the fragmentation are indeterminate. The resulting triple system is a hierarchical system, but may not be stable in the long term. The calculation could not be followed further due to the increasing number of particles contained in the fragments, lengthening the computational time.

In the isothermal case, using a minimum particle smoothing length, the outcome is a quadruple system. In this case, the discs around both initial protostars fragment at approximately $1.54 t_{\rm ff}$, well before periastron. The discs fragment earlier than in the polytropic equation of state case because there is less pressure support. With the lack of pressure support, the material must attain more of its support from rotation about the protostars, and the rotational instability increases. Also, once spiral arms are formed and collide, the resulting high-density region of gas is more prone to collapse under its self-gravity with an isothermal equation of state than it would be if it were heated. Again the calculation was halted due to increasing computational time at $2.2\ t_{\rm ff}$.

### 4.2 Evolution with sink particles

Although the use of a minimum particle smoothing length or a polytropic equation of state allows fragmentation calculations to be followed further than with a purely isothermal equation of state, the computational time still increases as more particles are accreted by the fragments, and soon becomes unacceptably long. Moreover, with the use of a minimum particle smoothing length, the internal fragment evolution is not followed accurately. Even with the polytropic equation of state, which provides a more physically correct evolution except in the very highest-density regions, the internal evolution of the fragments is not of interest. The main information to extract from fragmentation calculations is which fragments accrete the mass, and what the orbital parameters of the resulting stellar system are. By using sink particles in SPH, the internal evolution can be ignored, increasing the speed of the calculations, and allowing the final result of the fragmentation to be determined.

To ascertain if the use of sink particles in SPH affects the evolution of the cloud outside the fragments, and if so in what ways, the Boss & Bodenheimer (1979) test case was repeated using sink particles of varying accretion radii, with both isothermal and polytropic equations of state. The results from evolving the standard test case with sink particles using four different accretion radii and the polytropic equation of state are presented for comparison with the standard SPH calculation. The accretion radii were 0.02, 0.05, 0.10, and 0.20, in units of $10^{16}$ cm. The largest radius swallows almost the entire disc around each of the initial protostars. Since a larger fraction of the disc is neglected as the accretion radius increases, only the smallest-accretion-radius calculations might be expected to fragment the discs around the initial protostars. Also, interactions between protostars and discs will be affected in calculations where most of the disc is not modelled, due to the replacement of the disc's gravitational potential with a spherical potential.

Fig. 6 shows the results of a calculation using the polytropic equation of state and sink particles with boundary conditions and with $r_{\rm acc} = 0.20$. Two sink particles are dynamically created from the fragments at approximately $1.31 t_{\rm ff}$. Most of the volume of the discs that would form around these two sink particles is within the accretion radii, and hence not modelled. Only the very edges of the discs are still visible. Hence, no disc fragmentation occurs, and the result is a binary system. Overall, the dynamics of the calculation are similar to those for the calculation without sink particles. The calculation was followed to $4 t_{\rm ff}$. At this stage 79 per cent of the initial cloud is accreted by the sink particles. Most of the rest of the initial cloud material is expelled out to the radius of the initial cloud by gravitational torques from the binary.

**Figure 6.** The same calculation as presented in figure 5, but using dynamically created sink particles with accretion radii of $r_{\rm acc} = 0.20 \times 10^{16}$ cm. The sink particles are marked with crosses. Global evolution of the system is similar to figure 5, with the details of the calculation within the accretion radii of the sink particles being ignored. The use of sink particles enables the calculation to be followed much further than is possible with standard SPH. The simulation was performed with 8024 particles. The time, in units of the free-fall time ($t_{\rm ff} = 5.52 \times 10^{11}$ s), is given in the upper right hand corner of each panel. The axes are in units of $1.0 \times 10^{16}$ cm.

Fig. 7 gives the results of an identical calculation, but with $r_{\rm acc} = 0.02$. In this calculation, the majority of material in the discs around the protostars is left intact. The discs do not fragment. Again the evolution was followed to $4t_{\rm ff}$ and a binary system results with approximately 80 per cent of the initial cloud being contained by the sink particles, or their surrounding discs, with the rest being expelled out to the initial cloud radius. Both this calculation and the larger accretion radius calculation were easily followed further, up to $10t_{\rm ff}$, although, the constant-volume boundary conditions at the initial cloud radius gave unphysical results beyond about $4.5t_{\rm ff}$.

The dynamics of the gas in the discs in the small-accretion-radius calculations are similar, but not identical, to those of the calculation without sink particles. The evolution is not expected to be identical to the standard SPH calculation for three reasons. First, the gravitational potential of material inside an accretion radius is being approximated by a spherically symmetric distribution, although the error in this approximation should decrease as the accretion radius decreases. Secondly, the boundary conditions, although being a great improvement over no boundary conditions, are imperfect. The third difference, and perhaps the most important, is that, with the standard SPH calculations, the protostars are kept larger than is physically correct to allow the calculation to be followed. This results in the inner parts of a disc being replaced by a spheroidal protostellar object in approximate solid-body rotation due to the viscosity. This protostar helps to drive the disc fragmentation by becoming bar-unstable itself, as described in Section 4.1.

**Figure 7.** The same calculation as presented in figures 5 and 6, but using dynamically created sink particles with accretion radii of $r_{\rm acc} = 0.02 \times 10^{16}$ cm. The majority of the disc material for each protostar is outside the accretion radius and is still modelled by the calculation, with the result that the calculation follows the standard SPH calculation closer. The sink particles are not marked, but are contained within the central holes in the protostellar discs. The simulation was performed with 8024 particles. The time, in units of the free-fall time ($t_{\rm ff} = 5.52 \times 10^{11}$ s), is given in the upper right hand corner of each panel. The axes are in units of $1.0 \times 10^{16}$ cm.

With sink particles, this does not occur. There is no central object to aid the fragmentation, and the discs are able to remain in Keplerian rotation right up to the accretion radius. Because of these differences, it is not obvious whether the calculations with small-accretion-radius sink particles, or the standard SPH calculations give more accurate disc evolution. In particular, because the whole process of disc fragmentation is highly chaotic and is determined by the above differences, it is uncertain whether the discs fragment to form close binaries. These calculations show the discs to be highly unstable and likely to fragment, but higher resolution calculations are required to determine whether disc fragmentation actually occurs or not. Regardless, a wide binary is certainly formed.

The orbits of the each of the two initial fragments are given up to $3t_{\rm ff}$ for each calculation in Fig. 8. These were calculated by finding the centres of mass of the sink particles and their discs out to a radius of $0.30 \times 10^{16}$ cm from each of the initial two sink particles. They are compared with the orbits of the centres of mass of the initial two fragments and their discs formed from the calculation without sink particles. The centres of mass of the initial two fragments have been plotted to avoid the confusion of the fragmenting discs, which is not expected to be followed correctly with such large accretion radii in any case. The only sink particle calculation to give disc fragmentation was the calculation with $r_{\rm acc} = 0.05$, in which both discs fragmented, forming a quadruple system.

**Figure 8.** The orbits of the centres of mass of each initial fragment formed in the standard test case evolved using sink particles and a polytropic equation of state. The orbits are plotted from $1.3 t_{\rm ff}$ to $3.0 t_{\rm ff}$. The sink particles accretion radii are $r_{\rm acc} = 0.02$ (dotted), $r_{\rm acc} = 0.05$ (dashed), $r_{\rm acc} = 0.10$ (long-dashed), and $r_{\rm acc} = 0.20$ (dot-dashed). These orbits are compared to the orbits of the centres of mass, up to $2.2 t_{\rm ff}$, of the two initial fragments formed in the polytropic equation of state calculation without sink particles (solid line). The centres of mass of the fragments are calculated from all the particles within a radius of $0.30 \times 10^{16}$ cm from the initial sink particle (or most dense particle for the standard SPH calculation) of each fragment. It is noted that the discs around both the initial sink particles in the $r_{\rm acc} = 0.05$ calculation fragmented to form close binaries, and in this case both sink particles in each of the close binaries were used for the centre of mass calculation. The initial cloud free-fall time $t_{\rm ff} = 5.52 \times 10^{11}$ s. All distances are in units of $1.0 \times 10^{16}$ cm.

The orbits of the initial fragments in Fig. 8 are all similar until periastron. After periastron the orbits diverge, producing a scatter in the orbits. This is due to the different way each calculation models the interactions between the protostars and discs at periastron. In particular, after periastron, the largest-accretion-radius calculation, in which the protostars and discs are almost entirely contained inside the accretion radii, has a tighter orbit than the second-largest-accretion-radius calculation where parts of the discs are still modelled. For calculations that have parts of the discs modelled outside the accretion radius, the orbits tend to become tighter as the accretion radius decreases and more of the disc is modelled. Two mechanisms are presumably responsible for these effects. The first is the process of angular momentum transport via gravitational torques between the discs' spin angular momenta and the protostars' orbital angular momenta near periastron. The second is the loss of material from the outer edges of the discs which is expelled from the system during the periastron passage. These two mechanisms give a possible explanation for the results. With $r_{\rm acc} = 0.10$, the interaction, via gravitational torques, of the protostars with the discs, which are rotating more quickly, transfers angular momentum from the discs' spin angular momenta to the orbital angular momenta of the protostars. This gives a larger orbit than the case with $r_{\rm acc} = 0.20$ where there can be no such interaction. For the cases where the disc is modelled outside the accretion radius, the decreased apastron separation as the accretion radius is reduced may be due to greater loss of material from the edges of the discs from the system at periastron. The spiral arms passing through the discs act via gravitational torques to move angular momentum further out in the disc. With more of the disc modelled, the outer disc material may be expected to acquire more angular momentum from the inner parts of the discs, resulting in more material being expelled by the binary at periastron as the edges of the discs are disrupted. The loss of the orbital angular momentum of the system in expelling this material could give the tighter orbits with decreasing accretion radius seen in Fig. 8, since more material will be expelled. This mechanism may help to explain the observed absence of medium-period, high-eccentricity binary systems in stellar surveys (Duquennoy & Mayor 1991). These interactions between protostars and discs require further study.

In summary, for each calculation, the global results are the same. In all cases a binary system is formed with each component containing approximately 40 per cent of the original cloud mass at $4 t_{\rm ff}$. The remainder of the mass is expelled, by the binary, beyond the initial cloud radius. The orbits, until periastron, are essentially independent of the accretion radius used. However, at periastron approach, interactions between the protostars and the discs require that the discs be resolved if the evolution of the system is to be correctly followed. In addition, each of the binary components either consists of a protostar surrounded by a disc, or may itself consist of a binary, if the discs surrounding the initial protostars fragment. The use of sink particles allows the calculations to be followed to completion in a realistic amount of computational time. The polytropic equation of state calculation without sink particles requires 20 days of Sun Sparcstation 5 CPU time to evolve it to $2.2 t_{\rm ff}$, whereas with sink particles the calculations take between a half a day and two days to reach the same point, depending on the size of the accretion radius that is used. Moreover, the calculation time is constantly decreasing with sink particles due to the decreased number of active particles, whereas the time is constantly increasing without sink particles as more particles join the protostellar fragments. Hence, sink particle calculations can be evolved until all of the initial cloud material has been either accreted or expelled from the system.

Finally, the Boss & Bodenheimer (1979) sink particle calculations evolved with an isothermal equation of state allow disc fragmentation to occur far more easily than with the polytropic equation of state, and the discs around each of the initial protostars fragment early in the calculation even with the largest accretion radius of $r_{\rm acc} = 0.20$. It is clear that the initial result is still a binary system whose parameters are similar to those given by the polytropic equation of state calculations. However, each of the initial components of the binary itself consists of a closer binary and the separations of these closer binaries are often not small enough for the system to be a stable hierarchical system.

## 5 CONCLUSION

A new method for extending molecular cloud fragmentation calculations beyond protostar formation has been presented. It allows these calculations to be followed for as long as is required for all the initial cloud material to be either accreted by the protostars, or expelled from the system. This is achieved by using a smoothed particle hydrodynamics code where each protostellar fragment that forms during the collapse of a molecular cloud is replaced by a single, non-gaseous particle. This particle accretes infalling material within a specified radius, and is surrounded by appropriate boundary conditions to avoid the presence of the accretion hole adversely affecting the dynamics of material outside the accretion zone.

As a demonstration of this technique, the Boss & Bodenheimer (1979) standard isothermal test case for the fragmentation of a molecular cloud has been evolved until approximately 80 per cent of the initial cloud is contained in the protostars, or their surrounding discs, and the rest of the material has been expelled out to the radius of the initial cloud. It is found that a wide binary system is formed, with roughly equal mass components. Each component may itself consist of a close binary system due to the fragmentation of the discs that form around the two initial protostars, depending primarily on how the equation of state of the material is treated after it becomes optically thick.

Using this new technique, the gap between cloud fragmentation calculations and observations of stellar systems will be able to be bridged. This will finally allow comparison between the results of cloud fragmentation calculations and the observed characteristics of pre-main-sequence and main-sequence stellar systems.

## ACKNOWLEDGMENTS

We thank Jim Pringle, Cathie Clarke and Melvyn Davies for many useful discussions and a critical reading of the manuscript. We also thank the IoA system managers for their continuing efforts to provide the desired computational resources. MRB is grateful for a scholarship from the Cambridge Commonwealth Trust. IAB is grateful to PPARC for a postdoctoral fellowship. NMP is grateful to PPARC for financial support.